\documentstyle[epsf,prl,aps]{revtex}

\def\Pom{{\bf I\!P}}
\def\lsim{\mathrel{\rlap{\lower4pt\hbox{\hskip1pt$\sim$}}
    \raise1pt\hbox{$<$}}}         
\def\gsim{\mathrel{\rlap{\lower4pt\hbox{\hskip1pt$\sim$}}
    \raise1pt\hbox{$>$}}}         

\begin{document}

\twocolumn[\hsize\textwidth\columnwidth\hsize\csname@twocolumnfalse\endcsname

\title{Azimuthal asymmetry as a new handle on $\sigma_{L}/\sigma_{T}$ in diffractive DIS}

\author{ N.N.~Nikolaev$^{a,b}$, A.V.~Pronyaev$^{c}$, B.G.~Zakharov$^{b}$ }

\address{ $^{a}$IKP(Theorie), FZ J{\"u}lich, 5170 J{\"u}lich, Germany. \\
$^{b}$D. Landau Institute for Theoretical Physics, GSP-1, 117940, 
ul. Kosygina 2, Moscow V-334, Russia. \\
$^{c}$Virginia Polytechnic Institute and State University, Blacksburg, VA 24061-0435, USA. }

\maketitle

\begin{abstract}
We propose a new method of the determination of  $R^{D}=
\sigma_{L}^{D}/\sigma_{T}^{D}$ from the dependence of the diffractive cross 
section on the azimuthal angle between the electron scattering and 
proton scattering planes. The method is based on our finding of the 
model independence of the ratio of the $LT$ interference and transverse
diffractive structure functions. The predicted azimuthal asymmetry is 
substantial and can be measured at HERA. We show that the accuracy of our
reconstruction of $R^{D}$ is adequate for a reliable test of an 
important pQCD prediction of $R^{D}\gsim 1$ for large $\beta$.
\end{abstract}
\vskip1pc]

\narrowtext
The ratio of (L) longitudinal  and (T) transverse  cross sections, 
$R=\sigma_{L}/\sigma_{T}$, is a much discussed test of mechanisms 
of deep inelastic scattering (DIS). The QCD theory of diffractive DIS
(DDIS) 
$e p \rightarrow e'p'X$ predicts \cite{GNZlong} an unprecedented 
dominance of higher twist $\sigma_{L}^{D}$ over leading twist
$\sigma_{T}^{D}$ in a broad range of $\beta \gsim 0.9$ (Hereafter
the superscript $D$ stands for "diffractive" and the electron 
inelasticity $y$, the $\gamma^{*}p$ c.m.s. energy $W$, $Q^2$, 
$x=Q^2/(Q^2+W^2)$, the mass $M$ of the diffractive system $X$, 
$\beta =Q^2/(Q^2+M^2)$ and $x_{\Pom}=x/\beta$ are the standard DDIS
variables). 
This must be contrasted to leading twist $\sigma_{L}$ 
in small-$x$ inclusive DIS where the theory predicts \cite{NZHERA} 
$R=\sigma_{L}/\sigma_{T} \sim$ 0.2-0.3 in agreement with the 
experiment \cite{FLHERA}. The pQCD 
calculations of $\sigma_{L}^{D}$ describe the experimental data for 
large $\beta$ very well \cite{BGNPZ98} and higher twist $\sigma_{L}^{D}$
found in \cite{GNZlong} has become a part of modern 
parameterizations of DDIS structure functions (SF's) \cite{Bartels}. 
However, a direct measurement of $R^{D}$ requires a 
variable electron/proton energy runs which are not 
foreseen in the near future at HERA. Here below we show 
how this experimental limitation can be circumvented by the measurement 
of the LT interference SF $F_{LT}^{D}$. 

The major point behind our proposal \footnote{The preliminary results 
from this study have been reported elsewhere \cite{DIS97}} is that 
Leading Proton Spectrometers  of ZEUS and H1 opened an  
access to the transverse momentum $\vec{\Delta}$  of the recoil proton 
$p'$ and an azimuthal angle $\phi$ between the 
electron scattering and proton production planes. The incident 
$\gamma^{*}$ beam is a mixture of T and L photons
and an interference of diffractive transitions $\gamma^{*}_{T,L} 
\rightarrow X$ 
leads to the $\cos\phi$ dependence of the observed cross section. 
Incidentally, such an $LT$ interference is the s-channel helicity 
non-conserving (SCHNC) effect. There has been much discussion of different 
aspects of a related $LT$ interference in exclusive electroproduction 
\cite{SZ95} and quasielastic scattering $A(e,e'p)$ \cite{Boffi}. The
specific issue which we address in this paper is whether it is 
really possible to deduce $R^{D}={\sigma_{L}^{D}/\sigma_{T}^{D}}$
from the experimentally measured $\sigma_{2}^{D}=\sigma_{L}^D+\sigma_{T}^D 
\propto \sum_{X} \left\{ |A_{L}(\gamma^{*}\to X)|^{2}+ 
|A_{T}(\gamma^{*}\to X)|^{2}\right\}$ and $\sigma_{LT}^D \propto
{\rm Re} \sum_{X}  A_{L}^{*}(\gamma^{*}\to X)
A_{T}(\gamma^{*}\to X)$ and whether the 
so reconstructed $R^{D}$ will be model independent and under the 
control of perturbative QCD?

The principal point is as follows. For unpolarized electrons the 
differential cross-section of DDIS can be decomposed as 
\begin{eqnarray}
&&Q^{2}y{d\sigma(ep \rightarrow ep'X)\over dQ^2 dy dM^2 d\Delta^2 d\phi}
= {\alpha_{em}\over 2\pi^2}
\left[\left(1-y+{y^2\over 2}\right){d\sigma^{D}_{T} \over dM^2 d\Delta^2} 
\right.
\nonumber \\
&&+\left(1-y\right){d\sigma^{D}_{L} \over dM^2 d\Delta^2}
+\left(1-y\right){d\sigma^{D}_{TT} \over dM^2 d\Delta^2} \cdot\cos{2\phi}
\nonumber \\
&&\left.
+\left(1-{y\over 2}\right)\sqrt{1-y}\;
{d\sigma^{D}_{LT} \over dM^2 d\Delta^2} \cdot\cos{\phi}\right] =
\nonumber\\
&&{\alpha_{em}\over 2 \pi^{2}}\cdot
\left(1-y+{y^2\over 2}\right)\cdot{d\sigma^{D}_{2} \over dM^2 d\Delta^2}
\left\{1
-{y^{2}R^{D} \over 2(1-y)+y^{2}} +\right.
\nonumber\\
&&
\left.
{2(1-y)A_{TT}\over 2(1-y)+y^{2}}\cos 2\phi
+{(2-y)\sqrt{1-y} A_{LT}\over 2(1-y)+y^{2}}\cos \phi\right\}\, .
\label{eq:1}
\end{eqnarray}
The experimental measurement of the 
asymmetry $\propto \cos{\phi}$ amounts to the determination of 
\begin{equation}
A_{LT}={F_{LT}^{D} \over F_{T}^{D}+F_{L}^{D}}=
{\rho_{LT}\over 1+R^{D}}\, .
\label{eq:2}
\end{equation}
(for the sake of simplicity we focus on $y\ll 1$ which is the standard
experimental cutoff). If $\rho_{LT}=F_{LT}^{D}/F_{T}^{D}$ were known, 
then inversion of (\ref{eq:2}) would yield  
\begin{equation}
R^{D(4)}=\frac{\rho_{LT}}{A_{LT}}-1 \,. 
\label{eq:3}
\end{equation}
Below we argue for $\beta \gsim 0.9$ of the interest the theoretical
evaluations of $\rho_{LT}$ are model-independent and the inversion 
(\ref{eq:3}) offers a viable test of the pQCD prediction $R^{D}\gsim 1$ 
from the experimentally measured azimuthal asymmetry $A_{LT}$.  

The microscopic QCD mechanism of DDIS at $\beta >0.9$ is an excitation 
of $q\bar{q}$ Fock states of the photon (we focus on continuum 
$M^{2}\gg 4m_{f}^{2}$). The kinematical
variables are shown in fig.~1, quark and antiquark carry a fraction $z$ 
and $1-z$ of the photon's  momentum and 
\begin{equation}
M^{2}=Q^{2}{1-\beta \over \beta}={\vec{k}^{2}+ m_{f}^{2}\over z(1-z)}\, .
\label{eq:4}
\end{equation}
\begin{figure}
\begin{center}
\epsfysize 1.5 in 
\epsfbox{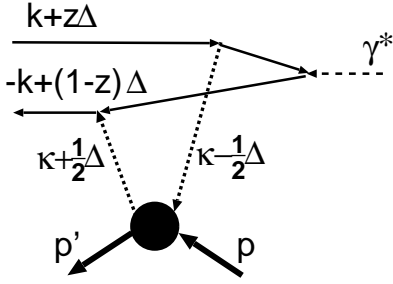}
\end{center}
\caption{
The sample Feynman diagram for diffraction excitation of the
quark-antiquark Fock state of the virtual photon.}
\label{fig1}
\end{figure}
\noindent
In our calculation of SCHNC $\sigma_{LT}^{D}$ which is a new result of this 
work we rely heavily upon the theory of the $s$-channel helicity conserving 
(SCHC) DDIS developed in \cite{GNZlong,BGNPZ98,NZsplit,GNZcharm}.
Extension \cite{NPZslope} of this formalism to the $\vec{\Delta}\neq 0$ 
gives ($i=L,T,LT,TT$, $n_{L}=n_{T}=0, n_{LT}=1,n_{TT}=2$)
\begin{eqnarray}
&&{d\sigma_{i}^{D} \over dM^2 d\Delta^2}\cdot \cos( n_{i}\phi)=
\nonumber\\
&&\frac{\alpha_{em}}{12\pi} \sum_{f} e_f^2 
\int d^2\vec{k}{k^{2}+m_{f}^{2} \over JM^{4}}\alpha_{S}^2(\overline{Q}^{2})
[H_i(z_{+})+H_i(z_{-})]\, ,
\label{eq:5}
\end{eqnarray}
where $e_{f}$ is the quark charge in units of the electron charge, $m_{f}$
is the quark mass, $z_{\pm}={1\over 2}(1\pm J),~ 
J^2=1-{4(k^2+m_f^2)\over M^2}$ and the hard scale 
$\overline{Q}^{2}$ is defined below. 
For the SCHC cross sections we have the familiar results
\cite{NZsplit} $H_{T}=[1-2z(1-z)]\vec{\Phi}_1^2+m_f^2 \Phi_2^2$ and
$H_{L}=4z^2(1-z)^2 Q^2 \Phi_2^2$, whereas for the SCHNC cross sections
we find 
\begin{eqnarray}
&&H_{LT}=4z(1-z)(1-2z)Q(\vec{\Phi}_1\vec{t})\Phi_2\, ,
\label{eq:6}\\
&&H_{TT'}=
2z(1-z)[\vec{\Phi}_1^2-2(\vec{\Phi}_1\vec{t})^2]\, ,
\label{eq:7}
\end{eqnarray}
where $\vec{t}$ is a unit vector tangential to the $(e,e')$ scattering plane 
and orthogonal to the $\gamma^{*}p$ collision axis.
Diffractive  amplitudes $\vec{\Phi}_{1}$ and $\Phi_{2}$ are calculable 
\cite{BGNPZ98,NZsplit,NPZslope} in terms of 
the lightcone wave function of the $q\bar{q}$ Fock state of the photon 
$\psi_{2}(z,\vec{k}) = {1/[\vec{k}^{2}+m_{f}^{2}+z(1-z)Q^{2}]}$ (we also 
use $\vec{\psi}_{1}(z,\vec{k})=\vec{k}\psi_{2}(z,\vec{k})$) and gluon density 
matrix of the proton  
${\cal{F}}\left(x_{\Pom},\vec{\kappa},\vec{\Delta}\right)$:
\begin{eqnarray}
&&\Phi_j={1\over 2\pi}\int {d^2\vec{\kappa} \over \kappa^{4}}
{\cal{F}}\left(x_{\Pom},\vec{\kappa},\vec{\Delta}\right)[\psi_j (z,\vec{r}+\vec{\kappa})
\nonumber\\
&&+\psi_j (z,\vec{r}-\vec{\kappa})-\psi_j (z,\vec{r}+{\vec{\Delta}\over 2})-
\psi_j (z,\vec{r}-{\vec{\Delta}\over 2})]\, ,
\label{eq:8}
\end{eqnarray}
and $\vec{r}=\vec{k}-{1\over 2}(1-2z)\vec{\Delta}$. The dependence of 
${\cal{F}}(x_{\Pom},\vec{\kappa},\vec{\Delta})$ on the variable 
$\vec{\Delta}\vec{\kappa}$ corresponds to the subleading BFKL singularities
\cite{Lipatov} and can be neglected for small $x_{\Pom}$ of the interest. 
For small $\vec{\Delta}^{2}$ within the diffraction cone
\begin{equation}
{\cal{F}}(x,\vec{\kappa},\vec{\Delta})=
{\partial G(x,\kappa^{2})\over \partial \log \kappa^{2}}
\exp(-{1\over 2}
B_{3\Pom}\vec{\Delta}^{2})\, ,
\label{eq:9}
\end{equation}
where  $\partial G/\partial \log \kappa^{2}$ is the conventional
unintegrated gluon structure function and
the diffraction cone $B_{3\Pom}\sim$ 6 GeV$^{-2}$ \cite{NNPZZ98}. 
As usual \cite{GNZlong,BGNPZ98,GNZcharm}, the dominant 
contribution to  $\vec{\Phi}_{1},\Phi_{2}$ comes from the 
leading log$\overline{Q}^{2}$ region of 
\begin{equation}
\vec{\kappa}^{2} \lsim \overline{Q}^{2} ={\vec{k}^{2}+m_{f}^{2} 
\over 1-\beta}
\label{eq:10}
\end{equation}
with the results
\begin{eqnarray}
&&\vec{\Phi}_{1}={2\vec{r}\over \overline{Q}^{4}} \left[ \beta + 
{m_{f}^{2}\over \overline{Q}^{2}}\right]G(x_{\Pom},\overline{Q}^{2})\, ,
\label{eq:11}
\\
&&\Phi_{2}={1\over \overline{Q}^{4}} \left[ 2\beta -1+
{2m_{f}^{2}\over \overline{Q}^{2}}\right]G(x_{\Pom},\overline{Q}^{2})\, .
\label{eq:12}
\end{eqnarray}

For small $\vec{\Delta}$ the nonvanishing  contribution to $\sigma_{LT}$ 
comes from the term 
$\propto (1-2z)\vec{\Delta}$ in $\vec{r}$, so that $H_{LT}
\propto (\vec{t}\vec{\Delta})=\Delta \cdot cos\phi$ as expected 
from the SCHNC interference of the 
helicity-flip and helicity
non-flip diffractive amplitudes. The typical contributions to the $k^{2}$ 
integrated cross section are proportional to integrals ($i=T,L,LT$)
\begin{eqnarray}
&&{\cal J}_{i}^{f}(x_{\Pom},
\langle\overline{Q}^{2}\rangle_{i})=(n+1)\int_0^{{M^2\over 4}-m_f^2} 
{dk^2 \over k^2+m_f^2}\times
\nonumber\\
&&\left[{m_f^2\over k^2+m_f^2}\right]^n
\left[\alpha_{S}(\overline{Q}^{2})G(x_{\Pom},\overline{Q}^{2})\right]^{2}
\label{eq:13}
\end{eqnarray}
where to the leading log $\langle\overline{Q}^{2}\rangle_{i}$ 
approximation and for heavy
flavours ${\cal J}_{i}^{f}= \left[\alpha_s(
\langle\overline{Q}^{2}\rangle_{i}) G(x_{\Pom},
\langle\overline{Q}^{2}\rangle_{i})
\right]^2$ \,, for the case of light flavours see a discussion in 
\cite{BGNPZ98}. We recall that for $\sigma_{T}^{D}$ which is twist-2 we 
have $n\geq 1$ and the dominant contribution to $\sigma_{T}^{D}$ comes 
from the aligned jet configurations, $k^2 \sim m_f^2$. Consequently 
\cite{BGNPZ98,GNZcharm} the relevant average QCD hard scale 
$\langle \overline{Q}^{2}\rangle_T \sim 
{m_{f}^{2} \over 1-\beta}$ which is always large for heavy flavours 
and/or $1-\beta \ll 1$. 

Our new finding is that $\sigma_{LT}$ also is dominated by the aligned 
jet configurations, which can readily be checked from eqs. 
(\ref{eq:5}),(\ref{eq:6}),(\ref{eq:11}) and (\ref{eq:12}).  
Consequently, we expect 
\begin{equation}
\langle\overline{Q}^{2}\rangle_{LT}\approx 
\langle\overline{Q}^{2}\rangle_{T}\, ,
\label{eq:14}
\end{equation} 
which is a basis for our conclusion on the model-independence of
$\rho_{LT}$. Indeed, our result for the SCHNC DDIS SF 
$F_{LT}^{D(4)}$ reads 
\begin{eqnarray}
&&F_{LT}^{D(4)}(\Delta^2, x_{\Pom}, \beta, Q^2)=
{\Delta\over Q}\cdot
\frac{2\pi}{9\sigma_{tot}^{pp}}\cdot
24\beta^4 (1-\beta)(2-3\beta)\times
\nonumber\\
&&\sum_{f}{ e_f^2 \over m_f^2}
{\cal J}_{LT}^{f}(x_{\Pom},\langle\overline{Q}^{2}\rangle_{LT})
\exp\left(-B_{LT}^{f}\Delta^{2}\right)\, .
\label{eq:15}
\end{eqnarray}
(Incidentally, $F_{LT}^{D}$ is twist-3). It must be compared to 
$F_{T}^{D(4)}$ of \cite{BGNPZ98}:
\begin{eqnarray}
&&F_{T}^{D(4)}(\Delta^2, x_{\Pom}, \beta, Q^2)=
\frac{2\pi }{9\sigma_{tot}^{pp}}\cdot
\beta(1-\beta)^2(3+4\beta+8\beta^2)\times
\nonumber\\
&&\sum_{f}{e_f^2\over m_f^2}
{\cal J}_{T}^{f}(x_{\Pom},\langle\overline{Q}^{2}\rangle_{T}) 
\exp\left(-B_{T}^{f}\Delta^{2}\right)\, .
\label{eq:16} 
\end{eqnarray}
Then, our finding of the proximity of the two hard scales (\ref{eq:14}) 
entails ${\cal J}_{LT}(x_{\Pom},\langle\overline{Q}^{2}\rangle_{LT})
\approx 
{\cal J}_{T}(x_{\Pom},\langle\overline{Q}^{2}\rangle_{T})$
and for the dominant contribution from light flavour excitation 
we obtain the fully analytic approximation 
\begin{eqnarray}
&&\rho_{LT}(\beta,\Delta)=\chi_{LT}(\beta)\cdot \frac{\Delta}{Q}\cdot
\exp(-[B_{LT}-B_{T}]
\vec{\Delta}^{2})=
\nonumber\\
&&\frac{24\beta^3\left(2-3\beta\right)}{\left(1-\beta\right)
\left(3+4\beta+8\beta^2\right)}\cdot \frac{\Delta}{Q}\cdot
\exp(-[B_{LT}-B_{T}]
\vec{\Delta}^{2})\, ,
\label{eq:17}
\end{eqnarray}
in which the {\sl r.h.s.} 
depends on neither $x_{\Pom}$ nor $Q^{2}$ (apart from the trivial 
kinematical factor $\Delta/Q$). For $\beta \gsim 0.9$ at practically
attainable $Q^{2} \lsim 100$ GeV$^{2}$ the open charm contribution to
$F_{T}^{D}$ and $F_{LT}^{D}$ is negligible small, see \cite{BGNPZ98}
and here below.

Obviously the accuracy of our evaluation of $R^{D}$ from 
eq.~(\ref{eq:3}) hinges on the knowledge of diffraction slopes $B_{i}$
($i=T,L,LT$) and an accuracy of the analytic approximation (\ref{eq:17}).
Given the parameterization of the gluon structure 
function of the proton, one can calculate the ratio $\rho_{LT}$ from 
the first principles using eqs. (\ref{eq:5})-(\ref{eq:8}) 
and compare such a numerical result with the analytic approximation 
(\ref{eq:17}). 

Before venturing into this numerical experiment, we notice that for
light flavours and moderately small $1-\beta \sim $ 0.05-0.1 the both 
$F_{LT}$ and $F_{T}$ receive certain contribution from the soft-to-hard 
transition region of $\overline{Q}^{2}$. For $Q^{2} \gsim Q_{c}^{2}$, 
where $Q_c^{2}(GRV)=0.4, Q_c^{2}(CTEQ)=0.49,Q_c^{2}(MRRS)= 1.25$ GeV$^{2}$,
we can evaluate ${\cal{F}}(x_{\Pom},Q^{2})$ using the standard 
GRV NLO \cite{GRV}, CTEQ 4LQ \cite{CTEQ} and/or MRRS ($m_c=1.5 GeV$) 
\cite{MRRS} parameterizations for the perturbative gluon SF of the 
proton and at soft $Q^{2}$ we apply to ${\cal{F}}(x_{\Pom},Q^{2})$ 
soft-to-hard interpolation described in \cite{BGNPZ98}. The minor
improvement over the related interpolation of $G(x,Q^{2})$ in 
\cite{BGNPZ98} is that we constrain the extrapolation of perturbative 
${\cal{F}}(x_{\Pom},Q^{2})$ to have the $Q^{2}$ dependence as 
in eq.~(14) of \cite{BGNPZ98} with the infrared cutoff parameter
$\mu_{G}=0.75$ GeV \cite{NZBFKL}. This way we correctly impose the 
gauge-invariance driven cancellation of radiation of soft gluons 
by colorless protons. Then we apply to ${\cal{F}}(x_{\Pom},Q^{2})$
the interpolation eq.~(15) of
\cite{BGNPZ98} and request the same numerical result for the
transverse diffractive SF $F_{T}^{D}$ at the typical $x_{\Pom}=10^{-3}$ 
and $Q^{2}=100$ GeV$^{2}$ and $\beta=0.7$ as found in \cite{BGNPZ98}, 
which fixes the nonperturbative soft contribution parameter: 
$C_{B}(GRV)=1.7, C_{B}(CTEQ)=1.86, C_{B}(MRRS)=2.0$. For all three 
models the unintegrated
gluon SF ${\cal{F}}(x_{\Pom},Q^{2})$ builds up from zero on a 
typical scale $\overline{Q}_{G}^{2}\sim 0.5$ GeV$^{-2}$ and  
for light flavours it is $\overline{Q}_{G}^{~-2}$ which supplants 
$m_{f}^{-2}$ as a scale in 
normalization of diffractive SF's (\ref{eq:15}),(\ref{eq:16}). 
We checked that when $\beta \to 1$ 
starting from this reference $\beta=0.7$ and the QCD scale $Q_{T}^{2}$ 
rises from the soft value to hard $Q_{T}^{2} \to {1\over 4}Q^{2}$, the 
values of $F_{T}^{D}$ for the three gluon densities diverge by $\lsim$ 
5-10$\%$ in conformity to the known slight divergence of the GRV, CTEQ 
and MRRS gluon densities. We checked also that the results for $F_{L}^{D}$ 
for the three models agree within the same accuracy. In order to not 
have very heavy figures, we usually show the numerical results for the 
CTEQ gluons. All the numerical results will be shown for $x_{\Pom}=10^{-3}$ 
and $Q^{2}=100$ GeV$^{2}$ which are typical of $\beta \sim 1$ kinematics 
at HERA. The continuum calculations can be trusted for $M^{2} \gsim 3$ 
GeV$^{2}$, beyond the exclusive resonance excitation, and we only 
consider $\beta <0.97$. 
\begin{figure}
\begin{center}
\epsfysize 2.0 in 
\epsfbox{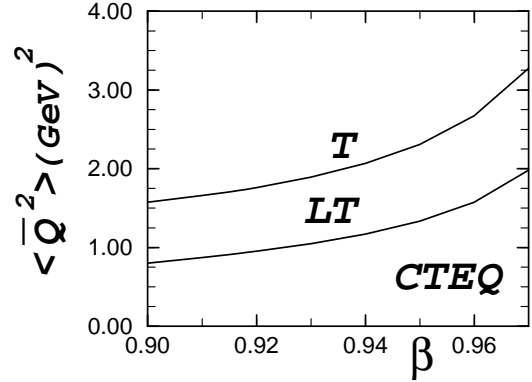}
\end{center}
\caption{
The average hard scale $\langle \overline{Q}^{2}\rangle$ for 
light flavour contribution to the 
($T$) transverse  and
$LT$ interference diffractive structure functions evaluated for
the CTEQ gluon SF of the proton.}
\label{fig2}
\end{figure}

The average value of the running hard scale (\ref{eq:10}) which enters
the calculations of the $\sigma_{LT}$ and $\sigma_{T}$ can be estimated
calculating the expectation values of $(\overline{Q}^{2})^{\gamma}$. 
For heavy flavours and weak scaling violations in the gluon structure 
SF these moments can be evaluated analytically with the result 
$\langle \overline{Q}^{2}\rangle_{T} 
= \left({6\over 5}\right)^{2}\langle \overline{Q}^{2}\rangle_{LT}$
for $\gamma={1\over 2}$,
the inequality being due to a somewhat different contribution of 
terms (13) with $n=1$ and $n=2$ in $\sigma_{LT}$ and $\sigma_{T}$. 
Numerical calculation for light flavours yields a comfortably large 
$\langle \overline{Q}^{2}\rangle = \langle (\overline{Q}^{2})^{\gamma}
\rangle^{1\over \gamma}$ shown in fig.~2 for the typical 
$\gamma={1\over 2}$ , which confirms the expectation (\ref{eq:14}).  
The slight inequality  $\langle \overline{Q}^{2}\rangle_{T} 
\approx 1.6 \langle \overline{Q}^{2}\rangle_{LT}$ is similar to that 
for heavy flavours and does not affect our major argument.

The calculation of the diffraction slope $B_{L,T}$ for the SCHC structure 
functions $F_{L,T}^{D}$ is found in \cite{NPZslope}, here we cite our 
new result for $B_{LT}$: 
\begin{eqnarray}
&&B_{LT}=B_{3\Pom}+
\nonumber\\
&&{1\over 20 m_f^2}\frac{(1-\beta)(2+7\beta+12\beta^2-483\beta^3+672\beta^4)}
{12\beta^2 (2-3\beta)}
\label{eq:18}
\end{eqnarray}
Here the second term is the contribution from the $\gamma^{*}X$ transition 
vertex and is a rigorous pQCD result because for heavy flavours $\langle \overline{Q}^{2}\rangle_{LT}$ 
is large for all $\beta$ (barring the vicinity of $\beta = {2\over 3}$ in 
which $F_{LT}^{D(4)}(\Delta=0)$ vanishes and $B_{LT}$ is ill defined). 
In the case of light flavours the scale of this 
component of $B_{LT}$ (and $B_{T}$ too \cite{NPZslope}) is rather set by 
$\overline{Q}_{G}^{~-2}$  than $m_{f}^{-2}$. The variations 
of diffraction slopes from CTEQ to 
GRV to MRS parameterizations do not exceed $\sim 5$ per cent and are not
shown in fig.~3. 
\begin{figure}
\begin{center}
\epsfysize 2.0 in 
\epsfbox{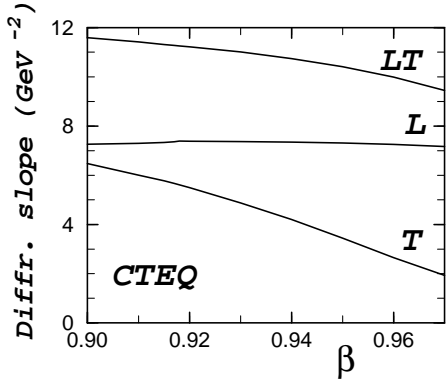}
\end{center}
\caption{
Our predictions for the diffraction slopes $B_{T}$, $B_{L}$, $B_{LT}$ of
the transverse, longitudinal and $LT$ interference diffractive SF's for 
the  CTEQ  gluon SF of the proton 
($x_{\Pom}=10^{-3}, ~ Q^{2}=100$ GeV$^{2}$).}
\label{fig3}
\end{figure}
 
\begin{figure}
\begin{center}
\epsfysize 2.0 in 
\epsfbox{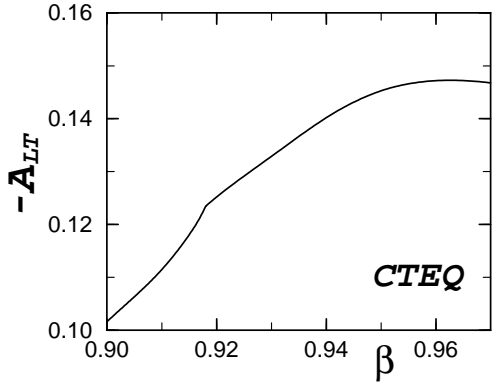}
\end{center}
\caption{
Our predictions for the $\beta$ dependence of the azimuthal asymmetry 
$A_{LT}$ for the $\Delta$-integrated cross sections evaluated using
the diffraction slopes of fig.~3 ($x_{\Pom}=10^{-3}, ~ Q^{2}=100$ GeV$^{2}$,
CTEQ gluon SF of the proton). 
}
\label{fig4}
\end{figure}

Evidently, a linear growth of azimuthal asymmetry
with $\Delta$ stops and the asymmetry reaches its maximal value at 
($B_{L}$ is more relevant in the region of the dominance of $F_{L}$)
\begin{equation}
\Delta^{2}={1 \over 2(B_{LT}-B_{L})} \sim {\rm (0.1-0.2)~ GeV^{2}}
\label{eq:19}
\end{equation}
which lies comfortably within the acceptance of the LPS's at HERA. 
With sufficiently high statistics, one can study experimentally the full
$\Delta$ dependence of the azimuthal asymmetry. In order to have the
first impressions of the expected signal, hereafter we shall consider 
the $\Delta$ integrated quantities, in which case
\begin{eqnarray}
&&\rho_{LT}={1\over 2}
\sqrt{{\pi \over B_{LT}Q^{2}}}\cdot {B_{T}\over B_{LT}}
\cdot\chi_{LT}(\beta)\, ,
\label{eq:20}
\\
&&R^{D} = R^{D}(\Delta=0){B_{T}\over B_{L}}\, .
\label{eq:21}
\end{eqnarray}

The predicted azimuthal asymmetry $A_{LT}$ is shown in fig.~4. It is 
substantial and within the reach of the LPS experiments at HERA. For 
$Q^{2}\sim 50$ GeV$^{2}$ for which $\beta \sim 0.95$ does still belong 
to the continuum, the expected asymmetry is $\approx 50\%$ larger. The 
cusp-like behaviour at $\beta=\beta_{c}=Q^{2}/(Q^{2}+4m_c^{2}) \approx 
0.92$ is due to the open charm excitation at $\beta < \beta_{c}$. 
\begin{figure}
\begin{center}
\epsfysize 2.0 in 
\epsfbox{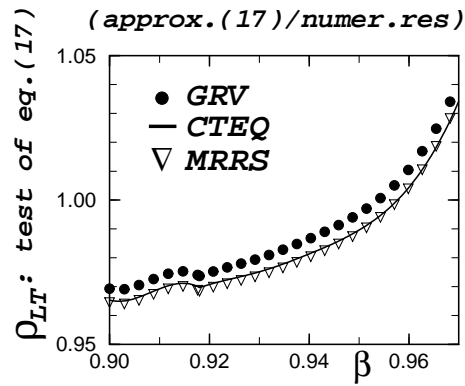}
\end{center}
\caption{
Numerical test of the accuracy of the analytic formula (\ref{eq:17}).
Shown is the ratio of approximation  (\ref{eq:17}) for $\rho_{LT}$ 
to the direct numerical evaluation of  $\rho_{LT}$ 
($x_{\Pom}=10^{-3}, ~ Q^{2}=100$ GeV$^{2}$).}
\label{fig5}
\end{figure}

Now we are in the position to test the accuracy of the analytic 
approximation (\ref{eq:17}). In fig.~5 we show the ratio of 
$\rho_{LT}$ given by (\ref{eq:17}) to the direct numerical
evaluation of $\rho_{LT}$.  A departure of their ratio from unity
does not exceed 5 per cent which confirms the expected high accuracy 
of our analytic result (\ref{eq:17}). The cusp is a weak effect of the 
open charm threshold which has been neglected in (\ref{eq:17}) but
included in the numerical evaluation of $\rho_{LT}$. The found variation 
from the GRV to CTEQ to MRRS gluons is negligible for the practical
purposes. 

The accuracy of reconstruction (\ref{eq:3}) of $R^{D}=\sigma_{L}^{D}/
\sigma_{T}^{D}$ can be judged from fig.~6a. Here we show the ratio 
of $R^{D}$  reconstructed from the numerically calculated azimuthal
asymmetry $A_{LT}$ on the basis of analytic approximation (\ref{eq:17}) for
$\rho_{LT}$ to the direct numerical result for $R^{D}$ presented in 
fig.~6b (for the convenience of plotting, in fig.~6b  we show the inverse 
quantity $1/R^{D}(\Delta=0$)). The reconstruction errors do dot
exceed dozen per cent and this accuracy of our reconstruction of 
$R^{D}$ is adequate for a reliable check of 
the pQCD predictions of large $R^{D}$ at $\beta \gsim 0.9$. We emphasize 
the $Q^{2}$ independence of $\langle\overline{Q}^{2}\rangle_{LT}$ and
$\langle\overline{Q}^{2}\rangle_{T}$  by which $\rho_{LT}$ does not
depend on $Q^{2}$. Consequently, our method allows a reliable 
experimental determination of the 
$Q^{2}$ dependence of $R^{D}$ and an unambiguous test of the higher
twist nature of $\sigma_{L}^{D}$.
\begin{figure}
\begin{center}
\epsfysize 3.5 in 
\epsfbox{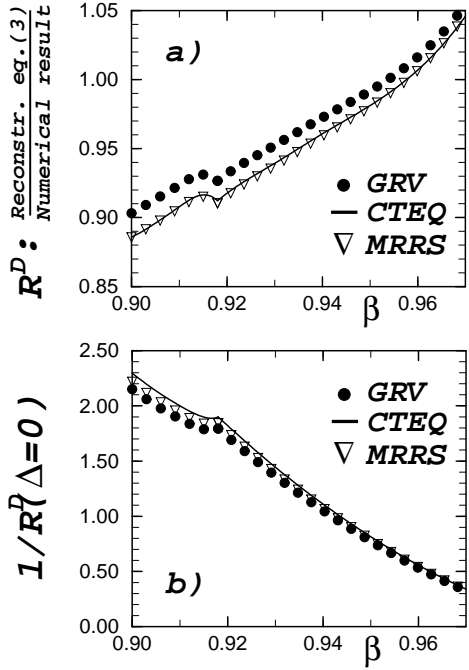}
\end{center}
\caption{
a) A comparison of the $R^{D}=\sigma_{L}/\sigma_{T}$ reconstructed 
from azimuthal asymmetry of fig.~4 using eqs.~(3) and approximation
(17) for $\rho_{LT}$ with the direct numerical evaluation of
$R^{D}$ for the GRV, CTEQ and MRRS gluon SF of the proton.
b) Our numerical predictions for $R^{D}(\Delta=0)$ in forward DDIS
for the same set of gluon SF's of the proton  
($x_{\Pom}=10^{-3}, ~ Q^{2}=100$ GeV$^{2}$).}
\label{fig6}
\end{figure}

To summarize, we derived the $s$-channel helicity nonconserving $LT$
interference diffractive structure function $F_{LT}^{D}$. It leads
to a substantial dependence of the diffractive cross section on the
azimuthal angle between the electron scattering and proton scattering
planes which can be observed at HERA. We demonstrated a weak model 
dependence of the ratio $F_{LT}^{D}/F_{T}^{D}$ for large $\beta$, which 
makes viable the reconstruction of the otherwise inaccessible 
$R^{D}=\sigma^{D}_{L}/\sigma^{D}_{T}$ from the experimentally measured 
azimuthal
asymmetry. We checked that the accuracy of our method is adequate for
the reliable test of the striking pQCD prediction of $R^{D}\gsim 1$ for
large $\beta$. 
 
The work of B.G.Z. has been supported partly by the INTAS grant 96-0597
and the work of A.V.P. was supported partly by the US DOE grant 
DE-FG02-96ER40994.

\end{document}